\documentclass[prl,reprint,showpacs,superscriptaddress]{revtex4-1}

\usepackage{graphicx}
\usepackage{textcomp}

\begin{document}


\title{Infinite Geometric Frustration in a Cubic Dipole Cluster} 



\author{Johannes Sch\"onke}
\affiliation{MPRG-ECPS, MPI f\"ur Dynamik und Selbstorganisation, Am Fa{\ss}berg 17, 37077 G\"ottingen, Germany}

\author{Tobias M. Schneider}
\affiliation{MPRG-ECPS, MPI f\"ur Dynamik und Selbstorganisation, Am Fa{\ss}berg 17, 37077 G\"ottingen, Germany}
\affiliation{Institute of Mechanical Engineering, EPFL, CH-1015 Lausanne, Switzerland}

\author{Ingo Rehberg}
\affiliation{Physikalisches Institut, Experimentalphysik V, Universit\"at Bayreuth, 95440 Bayreuth, Germany}


\date{\today}

\begin{abstract}
The geometric arrangement of interacting (magnetic) dipoles is a question of fundamental importance in physics, chemistry and engineering. Motivated by recent progress concerning the self-assembly of magnetic structures, the equilibrium orientation of 8 interacting dipoles in a cubic cluster is investigated in detail. Instead of discrete equilibria we find a new type of ground state consisting of infinitely many orientations. This continuum of energetically degenerate states represents a yet unknown form of magnetic frustration. The corresponding dipole rotations in the flat potential valley of this Goldstone mode enable the construction of frictionless magnetic couplings. Using novel computer-assisted algebraic geometry methods, we moreover completely enumerate all equilibrium configurations. The seemingly simple cubic system allows for exactly 9536 unstable discrete equilibria falling into 183 distinct energy families.
\end{abstract}



\maketitle 

Magnetism has fascinated mankind for millenia \cite{platon413bce}. Today, even the smallest magnets can hardly be overestimated in their relevance for magnetic storage technology. A fascinating example for the interplay of magnetic particles is their self-arrangement in cubic lattice clusters, see e.g. \cite{ahniyaz2007,ferrofluidposter2013}. Its macroscopic analogue is the toy known as ``magnetic cube puzzle'' shown in FIG.~1a, a stable arrangement of spherical magnets in a simple cubic cluster. How are these magnetic spheres oriented in such an ordered cluster? For the minimal arrangement within this class, namely a cube consisting of 8 magnets (see FIG.~1b), the answer is intriguing: There are infinitely many orientations! We find the ground state to be a continuum of energetically degenerate states -- an extreme form of magnetic frustration. The phenomenon of frustration arises when the system cannot simultaneously minimize all dipole-dipole interaction energies, see \cite{nisoli2013} for a recent review. As this continuum is the ground state of the cube system, the question arises: Are there any other equilibrium orientations? Through our novel application of methods from numerical algebraic geometry (see Supplementary~4) we are able to construct and classify the complete set of equilibrium states. This set comprises thousands of unstable discrete dipole orientations in addition to the continuous states. We stress here that we find \emph{all} equilibrium configurations (stable and unstable) unlike commonly used relaxation methods.

\begin{figure}[ht]
\centering
\includegraphics[width=\columnwidth]{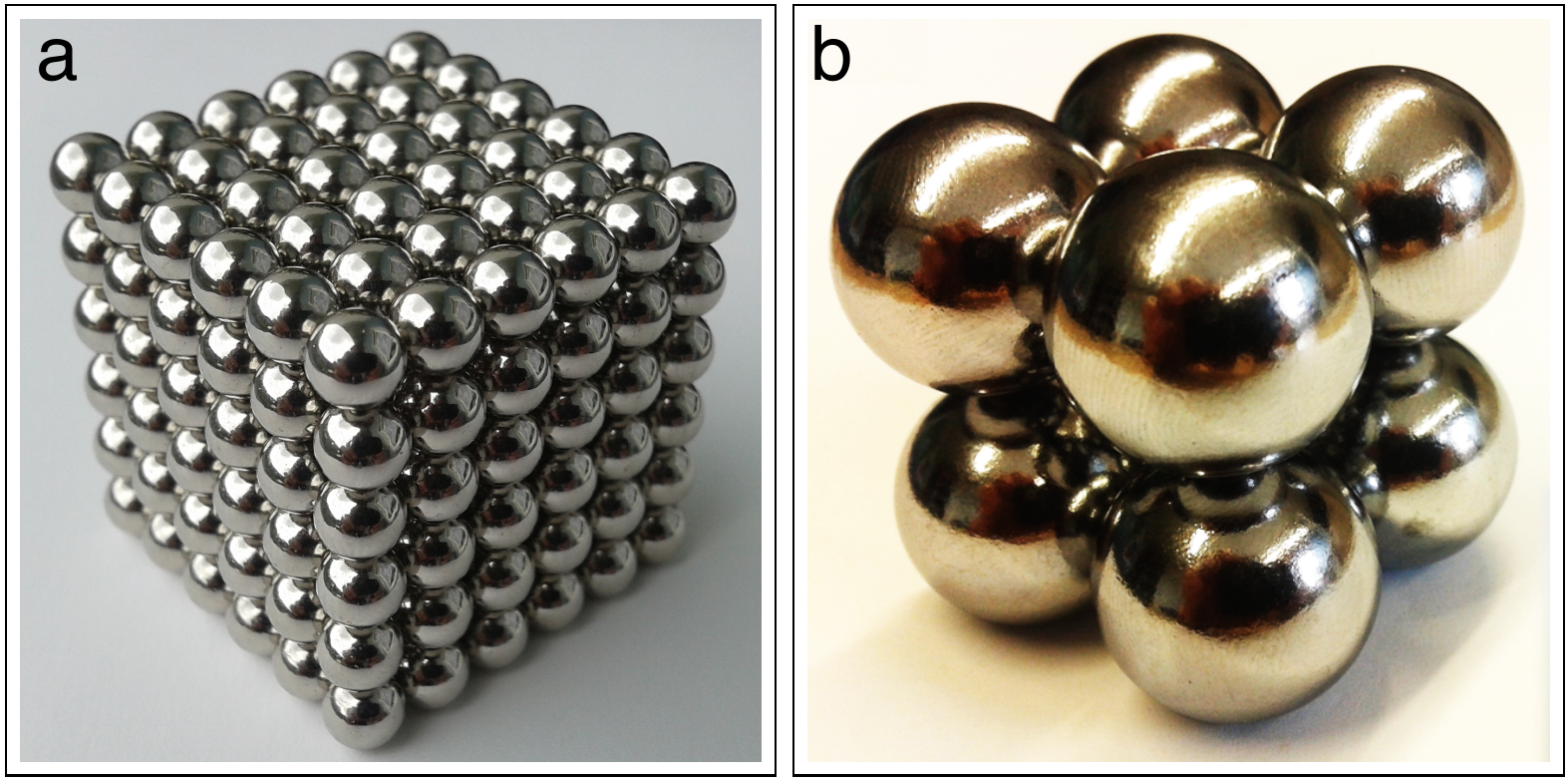}
\caption{(\textbf{a}) The ``magnetic cube puzzle'': A stable arrangement of 216 spherical magnets in a $6\times 6\times 6$ simple cubic cluster. (\textbf{b}) The minimal simple cubic configuration of spherical magnets, in a $2\times 2\times 2$ cluster. The ground state of this arrangement is not a single equilibrium configuration with only one discrete orientation for each dipole but rather a continuum of infinitely many configurations.}
\end{figure}

To our best knowledge continuous equilibrium states of dipole clusters have not been observed before. While dipole orientations have been investigated for planar quadratic configurations \cite{melenev2006} as well as for a uniform distribution on a sphere \cite{melenev2008}, those related works do not attempt to characterize the complete set of possible configurations nor do they study regular 3D clusters. In studies inspired by the self-assembly of magnetic spheres with variable orientation \emph{and} position, equilibrium configurations with a hierarchy of chains, rings and tubes have been found \cite{messina2014}.

\begin{figure*}[ht]
\centering
\includegraphics[width=\textwidth]{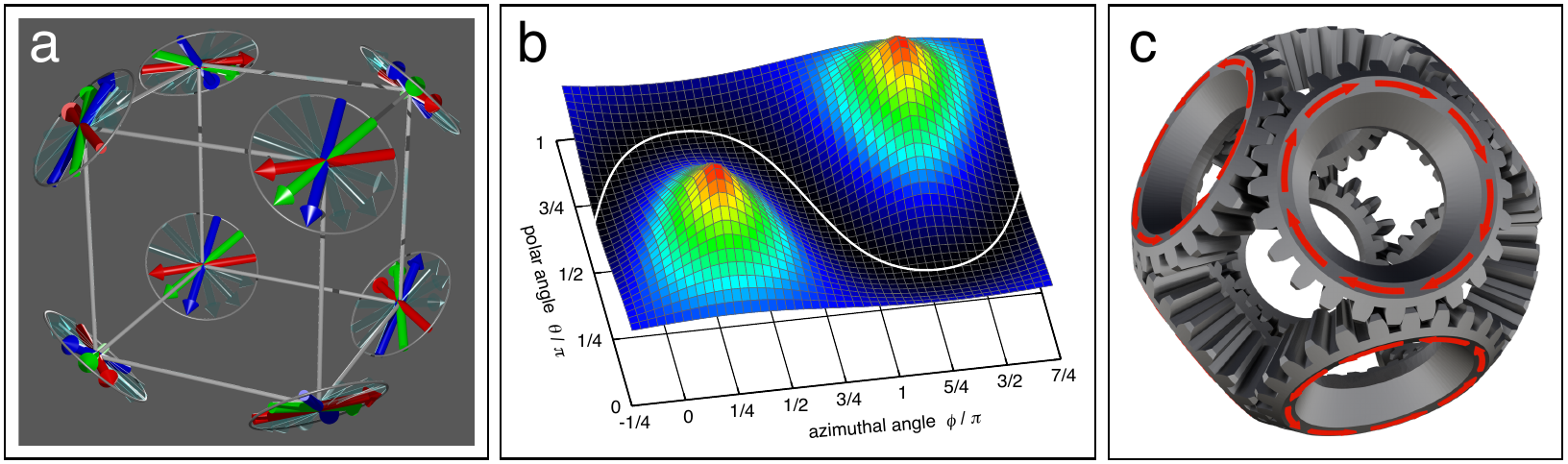}
\caption{(\textbf{a}) Rendering of 8 Dipoles as arrows located at the corners of a cube. The colored and translucent arrows show selected orientations in the ground state of the system which is a continuum of infinitely many configurations. The continuum corresponds to rotations of the dipoles in planes perpendicular to the cube's volume diagonal of the respective corner in analogy to the mechanical system in FIG.~2c. Consecutive arrows are seperated by {30\textdegree}. Along the continuum we repeatedly pass through two particularly notable configurations: Firstly, color red shows 2 counterrotating rings of 4 dipoles each lying in the upper and lower face of the cube. Secondly, color green shows 2 groups of 4 dipoles lying in 2 diagonal planes perpendicular to each other. Thirdly, color blue shows 2 rings similar to red but now in the left and right faces of the cube. Pairs of dipoles in opposing corners are always parallel. (\textbf{b}) Magnetic energy landscape for 8 dipoles located at the corners of a cube. The energy is shown as a function of the 2 orientation angles of one dipole, covering the range of all possible orientations. The other 7 dipoles adjust accordingly to the respective minimum energy configuration. The white line through the ``valley'' marks the ground state of the system which is a continuum of infinitely many configurations with identical energy (cf. FIG.~2a). (\textbf{c}) Rendering of 8 bevel gears located at the corners of a cube. The rotation axes point to the cube's center and every gear interlocks with its 3 edge neighbors. The motion of this mechanical system, indicated by the red arrows, corresponds to the motion of the ground state continuum in the magnetic system shown in FIG.~1b and 2a. This analogy opens up new engineering possibilities to construct frictionless magnetic couplings.}
\end{figure*}

The precise problem addressed in this paper reads: $N$ freely orientable dipoles of equal magnitudes are given together with their fixed positions in space. Which equilibrium configurations are possible? How many of these equilibria are stable? Which equilibrium represents the energetically favorable ground state, i.e. has the lowest energy? Here, we consider the classical dipole-dipole interaction with the magnetic energy
\[
E=\sum_{i<j}^N\frac{{\mathbf{m}}_i\cdot{\mathbf{m}}_j-3\,(\mathbf{m}_i\cdot\mathbf{e}_{ij})\,(\mathbf{m}_j\cdot\mathbf{e}_{ij})}{|\mathbf{r}_{ij}|^3}\,,
\]
where $\mathbf{m}_1,..,{\mathbf{m}}_N$ are the variable dipole moments with equal magnitudes $|\mathbf{m}_i|=1$, and $\mathbf{r}_{ij}$ denotes the fixed relative position vector between dipole $i$ and $j$ with $\mathbf{e}_{ij}$ being the corresponding unit vector, cf. Supplementary~1. The equilibrium condition $\nabla E=0$ corresponding to stationary points in the energy landscape is represented by a set of strongly coupled polynomial equations for all dipole orientations of the cluster, cf. Supplementary~2. Numerical algebraic geometry methods described in Supplementary~4 allow to construct the complete solution set and thereby find all equilibrium configurations.

\begin{figure*}[ht]
\centering
\includegraphics[width=0.92\textwidth]{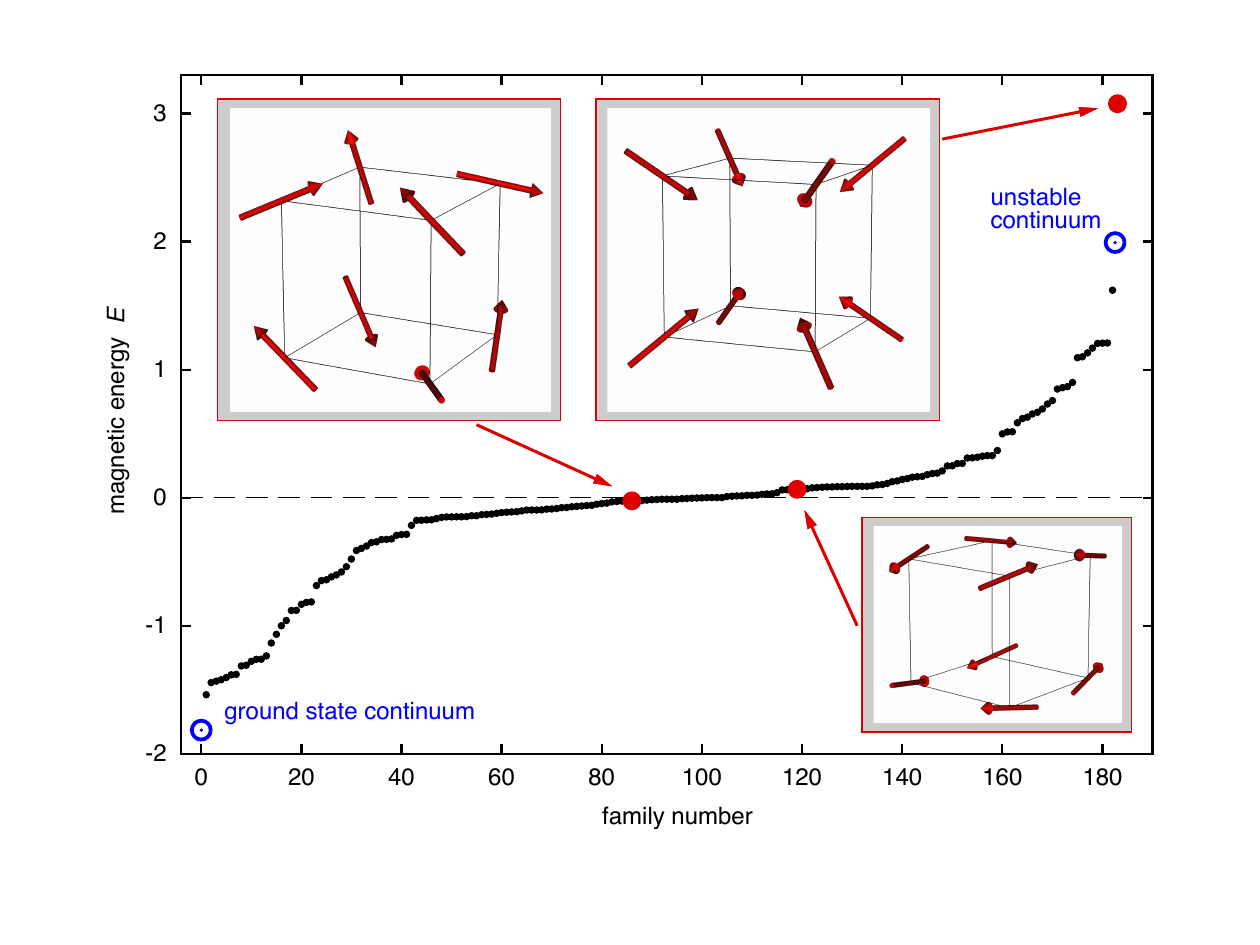}
\caption{Magnetic energy spectrum of all 183 families of discrete equilibrium configurations for 8 dipoles located at the corners of a cube. Each family (black filled circles) with its unique energy may contain up to 96 members due to polarity and cube symmetries. The red filled circles mark positions of exemplarily chosen configurations which are displayed in their respective inset. The upper left inset shows a fully unstructured configuration with no apparent symmetries, therefore the family has 96 members. The lower right inset shows a more structured configuration with some obvious symmetries, so this family has only 6 distinguishable members. The upper right inset shows the maximum energy configuration which is the most ordered one with all dipoles oriented to the cube center; all symmetries of the cube are retained and only the polarity flip gives a new configuration, so this family has 2 members. The additional 2 blue circles mark the energetic positions of the ground state continuum and the second real continuum.}
\end{figure*}

We investigate different elementary clusters of dipoles (see below) but report here in detail about one case which proves to be special: The case of 8 dipoles located at the corners of a cube. As pointed out before, the ground state of this arrangement is not a single configuration with only one discrete orientation for each dipole but rather a continuum of infinitely many configurations. Below we refer to this ground state continuum in short as the ``continuum'', its spatial structure is shown in FIG.~2a. The continuum exhibits a reflection symmetry through the 3 central planes (each parallel to a pair of cube faces) -- the dipole moments (as axial vectors) flip sign under reflection. If we rotate one of the dipoles along the continuum all the other dipoles rotate accordingly. Such a rotation is not affected by any magnetic counterforce since we stay on the same level in the magnetic energy landscape. This walk through the ``ground state valley'' is depicted in FIG.~2b. For the unit cube the energy of the continuum has the characteristic value of $E_{\sf c}=-2+\sqrt{2}/16+\sqrt{3}/18$ and its net magnetic moment is zero. Furthermore, the toroid moment with respect to the center of the cube is also zero, cf. Supplementary~1.

The continuum described here sheds a new light on frustration in magnetic systems, which has regained a lot of attention in recent years because of tailor-designed structures (e.g. ``artificial spin ice'') showing new and exciting thermodynamic behavior \cite{nisoli2013}. Inspired by the pioneering work of Pauling \cite{pauling1935} on water ice, theoretical investigations on Ising models \cite{wannier1950} and ``2D ice'' \cite{lieb1967}, upto the aforementioned artificial spin ices, frustration has been studied usually with the same basic assumptions: The system consists of spins with discrete (anti)ferromagnetic interactions between nearest neighbors, mostly on an infinite lattice, though some interesting studies on finite isolated magnetic clusters do also exist, e.g. \cite{mengotti2008}. In any of these spin systems frustration emerges from a \textit{countable} number of different states which are energetically degenerated. In contrast to spin systems we study classical dipoles which are freely orientable in space subject to dipole-dipole interactions. We consider not only nearest neighbor but fully coupled interactions in a finite system of $N$ dipoles. The resulting type of frustration in the cube continuum has a new quality: The finite system reported here has an \textit{uncountable} infinite number of different states which are energetically degenerated. In a sense, it is therefore ``infinitely frustrated''. Note that the continuous state is not a simple consequence of the individual dipoles being freely (continuously) orientable. In general the anisotropic dipole-dipole interaction induces discrete equilibrium configurations (see examples below) so that the cube continuum is indeed exceptional.

A further intriguing aspect of the continuum is the existence of an exact mechanical analogue. The aforementioned possibility to rotate one dipole along the continuum with the other dipoles following accordingly, raises the question: Can we reproduce the same dynamics through another type of interaction? It is possible with 8 bevel gears, as explained in FIG.~2c. The analogy between the mechanical and the magnetic system allows for new ways to engineer couplings. Since there are no magnetic counterforces to overcome along the continuum, one can build a frictionless sevenfold magnetic coupling. Depending on how close to the dipole approximation the actual magnets can be produced, we expect smooth performance. Classical gearbox damage is prevented through this contactless magnetic coupling.

The cube continuum is embedded in a richly structured state space which contains a multitude of other equilibrium configurations. We now provide the complete enumeration for all possible equilibria in the cube. This is a highly nontrivial problem since we aim at determining all solutions of a strongly coupled system of 32 polynomial equations (cf. Supplementary~2). The number of possible zero-dimensional (0D) solutions (corresponding to discrete isolated orientations) grows exponentially with the number of dipoles. For the cube we have a simple upper bound of $2^{24}=16,777,216$ possible 0D solutions, cf. Supplementary~3 for the derivation. Additionally, there may be higher dimensional solution manifolds, the continuum described above is a 1D example. Fortunately, this system size can be tackled with methods known under the term ``numerical algebraic geometry'' which were developed in the last two decades \cite{sommese2000,hauenstein2011,bertini2013}, Supplementary~4. The result is an astonishing number of 1,594,032 (generally complex) 0D solutions! Besides that there are four 1D continua, two of them being complex, plus the ground state continuum described above and a second real continuum. Higher dimensional solution manifolds do not exist (cf. Supplementary~4).

Extracting the physically meaningful real-valued subset of 0D solutions, we still end up with 9536 solutions. These can now be sorted into energy families, i.e. all solutions with identical energies belong to the same family. There are always at least 2 configurations with identical energy because the polarity symmetry (reversing the orientation of all dipoles) is again a solution and leaves the energy unchanged. In addition, there is the full symmetry group of the cube of order 48. Depending on the symmetries of the respective solution, we therefore may have up to $2\cdot 48=96$ members in one energy family. In general, there could be more members in a family if two configurations which are not related through symmetries have accidentally the same energy, although this does not happen in the cube. The sorting gives rise to 183 families of 0D solutions. The energy spectrum of these families together with some exemplary configurations is shown in FIG.~3.

The stability of any equilibrium configuration in our system is determined by the $2N$ eigenvalues $\lambda_k$ of the Hessian matrix $\mathbf H$, i.e. the matrix of all second order partial derivatives of the energy $E$ with respect to the $2N$ degrees of freedom. A general result for systems considered here is the relation
\[
\sum_{k=1}^{2N}\lambda_k=\mathsf{Tr}\,(\mathbf H)=-4E\,,
\]
see Supplementary~5 for the derivation. It shows that a positive energy $E>0$ is a sufficient condition for instability. Because then, the sum of all eigenvalues is negative, so there must be at least one negative $\lambda_k$, which classifies a configuration as unstable. From this we can conclude that a negative energy $E<0$ is a necessary condition for stability. For the cube this means that the second real continuum is unstable (cf. FIG.~3). Calculating the eigenvalues, we actually find all 183 families of discrete equilibria to be (unstable) saddles in the energy landscape, i.e. mixed positive and negative $\lambda_k$. The only exception is the maximum energy family which is necessarily unstable in any direction. This confirms that the ground state continuum is the only stable state.
\begin{table}[ht]
\begin{tabular*}{\hsize}{@{\extracolsep{\fill}} l c r r r r}
\hline\hline
Arrangement & $N$ & $S_{\sf max}=2^{3N}$ & $S$ & $S_{\sf real}$ & $F_{\sf real}$\\
    \hline
    Line segment&  2 &                64 &           8 &      8 &   4 \\ 
    Triangle    &  3 &               512 &          96 &     48 &   8 \\ 
    Tetrahedron &  4 &            4\,096 &         420 &    116 &  10 \\ 
    Octahedron  &  6 &          262\,144 &     37\,608 & 1\,156 &  43 \\ 
    Cube        &  8 &      16\,777\,216 & 1\,593\,776 & 9\,536 & 183 \\ 
    Icosahedron & 12 & 68\,719\,476\,736 &           ? &      ? &   ? \\
	\hline\hline
\end{tabular*}
\caption{Number of discrete equilibrium configurations (DEC) for different arrangements of freely orientable dipoles. The dipoles are positioned at the corners of the respective arrangement. $N$ is the number of dipoles, $S_{\sf max}$ is a simple upper bound (cf. Supplementary~3) for the number of possible DEC's, $S$ is the actual number of (generally complex) DEC's, $S_{\sf real}$ is the number of real-valued DEC's and $F_{\sf real}$ is the number of energy families the real-valued DEC's split into. The last row serves solely as an illustration of exponential complexity.}
\end{table}
Now we put the cube and its continuum into the context of other regular dipole clusters. We consider the dipoles to be located at the vertices of various regular geometric shapes. Figure 5 lists the number of solutions and energy families for different arrangements. The ground state continuum of the cube seems to be an exceptional property. So far we did not find any other regular arrangement which has this feature. Simple planar arrangements like the line segment or the equilateral triangle do not have continua, their ground states are necessarily discrete. The two smaller (in terms of numbers of corners) Platonic solids, i.e. tetrahedron and octahedron, have continua, but these are unstable. Therefore, their ground states are also discrete. Another common feature of the regular arrangements investigated (cf. TABLE~I) is the existence of only one stable configuration (modulo energetic degeneracies due to symmetries). For larger clusters, we expect several stable configurations to coexist.

In this study we report on a yet unknown type of ground state for systems of interacting dipoles -- a continuum of infinitely many energetically degenerate orientations. This result raises several new questions: Is the cube the only cluster that admits a stable (possibly ground state) continuum, allowing any number of dipoles in any arrangement? What happens to a continuum in an external field? What is the susceptibility of an arrangement possessing a continuum? What consequences do (stable or unstable) continua have for the dynamics of magnetic clusters, or more general, for their thermodynamic properties? This last question is especially important for the miniaturization of domains in magnetic information storage: The height of energetic barriers between different coding states limits the thermodynamic long-term stability. In our case the completely vanishing energy barrier of the continuum even prevents any information storage.


%
%

%

\begin{acknowledgments}
The authors are grateful to Michael Grunwald \cite{grunwald2014} for the design and construction of a dipole cube and for the rendering of FIG.~2c. JS thanks Hecke Schrobsdorff for an introduction to POV-Ray and Priya Subramanian for numerous fruitful distractions.
\end{acknowledgments}

\bibliography{references}

\end{document}


\baselineskip24pt


\section*{Supplementary Information}

\subsection*{Supplementary 1 - Energy, net moment and toroid moment}

The magnetic energy $U$ can be scaled with
\begin{equation*}
U_0=\frac{\mu_0m_0^2}{4\pi d^3}\,,
\end{equation*}
where $d$ is a characteristic lengthscale (e.g. the edge length of the cube), $m_0$ a dipole moment scale and $\mu_0$ the vacuum permeability. The dimensionless energy $E$ can then be defined as
\begin{equation}
E=\frac{U}{U_0}=\sum_{i<j}^N\frac{{\mathbf{m}}_i\cdot{\mathbf{m}}_j-3\,(\mathbf{m}_i\cdot\mathbf{e}_{ij})\,(\mathbf{m}_j\cdot\mathbf{e}_{ij})}{|\mathbf{r}_{ij}|^3}\,,
\label{eq_energy}
\end{equation}
where $\mathbf{m}_1,..,{\mathbf{m}}_N$ describe the dimensionless moments of the $N$ freely orientable dipoles of equal magnitude $|\mathbf{m}_i|=1$ and $\mathbf{r}_{ij}$ is the dimensionless relative position vector between dipole $i$ and $j$ with $\mathbf{e}_{ij}$ denoting the corresponding unit vector.

The net magnetic moment is defined as
\begin{equation*}
\mathbf{M}=\frac{1}{N}\sum_{i=1}^N\mathbf{m}_i\,,
\end{equation*}
which is a discrete analogue to the magnetization. The normalization of the moments $|\mathbf{m}_i|=1$ guarantees that $|\mathbf{M}|\leq 1$. The toroid moment is defined with respect to a given point $c$ in space as
\begin{equation*}
\mathbf{T}=\frac{1}{N}\sum_{i=1}^N\mathbf{p}_i\times\mathbf{m}_i\,,
\end{equation*}
where $\mathbf{p}_i$ is the vector from the point $c$ to the position of dipole $i$, see e.g. \cite{melenev2006}. The magnetic dipole moment $\mathbf{m}$ can be thought of as a current loop in the limit where its area $A$ goes to zero while its current $I$ diverges, keeping the product $|\mathbf{m}|=AI$ constant. In the same sense, the toroid moment $\mathbf{T}$ can be envisioned as a toroidal inductor coil in the limit where the torus radius $R$ and cross-sectional tube area $A$ go to zero while the current $I$ diverges, keeping the product $|\mathbf{T}|=RAI$ constant. $\mathbf{T}$ is useful to describe vortex-like states of dipole configurations, like e.g. planar head-to-tail ring configurations, for which $|\mathbf{T}|$ is maximal. For the cube continuum we have $\mathbf{M}=0$ and $\mathbf{T}=0$ with respect to the center of the cube.


\subsection*{Supplementary 2 - System of equilibrium equations}

Finding all equilibria of a dynamical system with energy $E$ is equivalent to finding all states of the system where the gradient vanishes. In general, the gradient has as many components as there are degrees of freedom in the system, in our case $2N$: Two angles per dipole to describe its orientation. However, we found that the use of cartesian coordinates to describe the dipole orientations is advantageous compared to the angle formulation: Firstly, we avoid the inherent coordinate singularities of spherical coordinates. Secondly, the degree of the resulting system of polynomial equations is lower (quadratic instead of quartic). Thirdly, the structure of the equations is highly symmetric.

Let the orientation of dipole $i$ be given as $\mathbf{m}_i=(x_i, y_i, z_i)^{\sf T}$, then we can define the full orientation vector 
\begin{equation*}
\boldsymbol{\Omega}\mathrel{\mathop:}=(x_1,y_1,z_1,...,x_N,y_N,z_N)^{\sf T}\,,
\end{equation*}
which contains the orientations of all dipoles. The energy $E$ can now be written as a symmetric bilinear form $E(\boldsymbol{\Omega},\boldsymbol{\Omega})$ with the representation
\begin{equation*}
E=\frac{1}{2}\,\boldsymbol{\Omega}^{\sf T}\mathbf{W}\,\boldsymbol{\Omega}
\qquad\mbox{or}\qquad
E=\frac{1}{2}\,\sum_{i,j=1}^{3N}\Omega^i\,W_i^j\,\Omega_j\,,
\end{equation*}
where $\mathbf{W}$ is the symmetric $3N\times 3N$ interaction matrix which encodes the positional information of all dipoles. $\mathbf{W}$ is constant for a given arrangement (like e.g. the cube). With the definition of the linear combinations
\begin{equation*}
L_{ik}\mathrel{\mathop:}=\sum_{j=1}^{3N}W_{3(i-1)+k}^j\,\Omega_j\,,\qquad i=1,2,...,N\,,\qquad k=1,2,3\,,
\end{equation*}
the zero gradient equations (i.e. equilibrium conditions) for all dipoles can be expressed in the form of the $3N$ cyclic equations
\begin{equation}
\begin{aligned}
L_{i1}\,y_i\;\; &=& L_{i2}\,x_i&\\
L_{i2}\,z_i\;\; &=& L_{i3}\,y_i&\qquad\qquad i=1,2,...,N\\
L_{i3}\,x_i\;\; &=& L_{i1}\,z_i&\,.
\end{aligned}
\label{eq_gradients}
\end{equation}
Since we assume fixed magnitudes for the moments $|\mathbf{m}_i|=1$ the cartesian description demands $N$ additional ``sphere'' equations
\begin{equation}
x_i^2+y_i^2+z_i^2=1\,,\qquad i=1,2,...,N\,.
\label{eq_spheres}
\end{equation}
Altogether Eqs. \eqref{eq_gradients} and \eqref{eq_spheres} set up a strongly coupled system of $4N$ quadratic polynomial equations in the $3N$ unknown components of $\boldsymbol{\Omega}$. 


\subsection*{Supplementary 3 - Upper bound for the number of isolated equilibria}

To get an upper bound for the number of possible isolated solutions, our overdetermined system Eqs. \eqref{eq_gradients} and \eqref{eq_spheres} with $4N$ equations for $3N$ unknowns has to be reformulated. We ``randomize down'' to a square system: The system is replaced by $3N$ random linear combinations of the original $4N$ equations, cf. \cite{sommese2000}. Then, a variant of Bertini's theorem (see e.g. \cite{bertini2013} p.163) assures that the set of isolated solutions for the randomized system is a superset of the original system. Therefore, every upper bound (for the number of isolated solutions) for the randomized system is one for the original system as well.

Now we use the basic version of B\'ezout's theorem: A polynomial system with $n$ variables and $n$ equations of total degree $d_1,..,d_n$ has at most $S_{\sf max}\mathrel{\mathop:}=d_1d_2...d_n$ isolated solutions. The randomized system with $3N$ variables consists of $3N$ quadratic polynomial equations (i.e. the total degree is always 2). Therefore we have $S_{\sf max}=2^{3N}$ possible isolated solutions, which is also an upper bound for the original system (see above). Compared to the analogous analysis for the angle formulation ($S_{\sf max}=2^{6N}$) this is a significantly better bound.


\subsection*{Supplementary 4 - Numerical solution method}

The system of polynomial Eqs. \eqref{eq_gradients} and \eqref{eq_spheres} is solved with the open-source software Bertini{\texttrademark}: Software for Numerical Algebraic Geometry \cite{bertini_homepage}. This numerical software framework is developed for industrial problems including mechanism and robot kinematics, chemistry, and computer-aided design, to name a few. However, it can be used to solve any system of polynomial equations. The manageable system sizes for numerical programs like Bertini{\texttrademark} are much greater compared to symbolic manipulation programs which utilize Gr\"obner bases and related methods, see e.g. \cite{singular_homepage}.

The numerical algorithm in Bertini{\texttrademark} is based on homotopy continuation and uses modern methods like randomization techniques and regenerative cascades, see \cite{hauenstein2011,bertini2013}. The general idea of homotopy continuation in order to solve a  system $\Sigma$, is to consider another system $\Sigma^\star$ with known solutions and deform $\Sigma^\star$ to $\Sigma$. The deformation is achieved through a parameterization of the system equations, such that they reduce to $\Sigma$ and $\Sigma^\star$ for different parameter values.

A parallel version of the software is available based on the Message Passing Interface (MPI). An overview of the theoretical background in algebraic geometry, a detailed description of the numerical algorithm and a comprehensive user manual for Bertini{\texttrademark} can be found in book \cite{bertini2013}.

The software provides information about the number of solution components and their dimensions. Further, for 0D (isolated) components it gives the numerical values of the solutions up to any required precision. For higher dimensional components any number of sampling points can be generated.


\subsection*{Supplementary 5 - Stability}

We now prove the formula connecting the trace of the Hessian matrix $\mathbf H$ (the sum of all eigenvalues $\lambda_k$) with the magnetic energy $E$
\begin{equation}
\sum_{k=1}^{2N}\lambda_k=\mathsf{Tr}\,(\mathbf H)=-4E.
\label{eq_trace}
\end{equation}
Here it is convenient to use spherical coordinates with the polar and azimuthal angles $\theta_i$ and $\phi_i$ to describe the orientation of dipole $i$
\begin{equation}
\mathbf{m}_i=\left(
\begin{array}{c}
x_i\\
y_i\\
z_i
\end{array}
\right)
=
\left(
\begin{array}{c}
\cos\phi_i\sin\theta_i\\
\sin\phi_i\sin\theta_i\\
\cos\theta_i
\end{array}
\right)\,.
\label{eq_spherical}
\end{equation}
The energy $E$ in Eq. \eqref{eq_energy} consists of sums of component-wise products of the dipole moments. If we define $\epsilon_i$ to collect all terms of $E$ containing $\theta_i$ and $\phi_i$ we can write with Eq. \eqref{eq_spherical}
\begin{equation}
\epsilon_i\mathrel{\mathop:}=A_i\cos\phi_i\sin\theta_i+B_i\sin\phi_i\sin\theta_i+C_i\cos\theta_i\,,
\label{eq_epsilon}
\end{equation}
where $A_i,B_i,C_i$ depend on all other angles but not on $\theta_i$ and $\phi_i$. Note that since all terms of $E$ consist of pairs of different moments, every term appears in two different $\epsilon_i$ and we have
\begin{equation}
E=\frac{1}{2}\sum_{i=1}^N\epsilon_i\,.
\label{eq_e_eps}
\end{equation}
We need the $2N$ diagonal entries of $\mathbf H$, which are the second covariant derivatives of $E$ with respect to all angles $\theta_1,\phi_1,\theta_2,\phi_2,...,\theta_N,\phi_N$. We may replace $E$ with the respective $\epsilon_i$ when differentiating with respect to $\theta_i$ and $\phi_i$ and get the diagonal entries
\begin{equation}
\begin{aligned}
H_{11}&=&\frac{\partial^2\epsilon_1}{\partial\theta_1^2}\,,\qquad
H_{22}&=&\frac{1}{\sin^2\theta_1}\frac{\partial^2\epsilon_1}{\partial\phi_1^2}+\cot\frac{\partial\epsilon_1}{\partial\theta_1}\,,\\
H_{33}&=&\frac{\partial^2\epsilon_2}{\partial\theta_2^2}\,,\qquad
H_{44}&=&\frac{1}{\sin^2\theta_2}\frac{\partial^2\epsilon_2}{\partial\phi_2^2}+\cot\frac{\partial\epsilon_2}{\partial\theta_2}\,,\\
&\vdots&&\vdots&
\end{aligned}
\label{eq_hessian}
\end{equation}
The extra ``$\cot\dots$'' term when differentiating with respect to $\phi$ is a direct result of the second covariant derivative in spherical coordinates. Inserting Eq. \eqref{eq_epsilon} into \eqref{eq_hessian} we find
\begin{equation*}
H_{11}=H_{22}=-\epsilon_1\,,\qquad
H_{33}=H_{44}=-\epsilon_2\,,\qquad
\dots
\end{equation*}
and therefore
\begin{equation*}
\mathsf{Tr}\,(\mathbf H)=\sum_{k=1}^{2N}H_{kk}=-2\sum_{i=1}^{N}\epsilon_i\,.
\end{equation*}
This together with Eq. \eqref{eq_e_eps} proves relation \eqref{eq_trace}. Note that the only condition on $E$ for Eq. \eqref{eq_trace} to hold is that $E$ consists only of component-wise products of two moments. There may be other types of interactions to which this relation is applicable.


\bibliographystyle{pnas}
\bibliography{references}
